\documentclass{article}
\usepackage{graphicx}
\usepackage[flushleft]{threeparttable}
\graphicspath{ {./images/} }
\usepackage{tocloft}  
\usepackage{titletoc} 
\usepackage{etoolbox} 
\usepackage{hyperref}

\usepackage[english]{babel}

\usepackage{listings}
\usepackage{xcolor}
\usepackage{makecell}
\usepackage{url}

\lstdefinestyle{mystyle}{
  backgroundcolor=\color{backcolour},   commentstyle=\color{codegreen},
  keywordstyle=\color{magenta},
  numberstyle=\tiny\color{codegray},
  stringstyle=\color{codepurple},
  basicstyle=\ttfamily\footnotesize,
  breakatwhitespace=false,         
  breaklines=true,                 
  captionpos=b,                    
  keepspaces=true,                 
  numbers=left,                    
  numbersep=5pt,                  
  showspaces=false,                
  showstringspaces=false,
  showtabs=false,                  
  tabsize=2
}

\usepackage[a4paper]{geometry}

\usepackage{amsmath}
\usepackage{graphicx}
\usepackage{multirow}
\usepackage{authblk}

\usepackage[version=4]{mhchem}
\usepackage{booktabs}
\usepackage{csquotes}
\usepackage{setspace}
\title{A predictive machine learning force field framework for liquid electrolyte development}

\author[1,*]{Sheng Gong}
\author[1,*]{Yumin Zhang}
\author[2,*]{Zhenliang Mu}
\author[2]{Zhichen Pu}
\author[2,+]{Hongyi Wang}
\author[2]{Xu Han}
\author[1]{Zhiao Yu}
\author[2,+]{Mengyi Chen}
\author[2]{Tianze Zheng}
\author[1]{Zhi Wang}
\author[2]{Lifei Chen}
\author[1]{Zhenze Yang}
\author[1]{Xiaojie Wu}
\author[2]{Shaochen Shi}
\author[2,**]{Weihao Gao}
\author[1,**]{Wen Yan}
\author[2]{Liang Xiang}

\affil[1]{ByteDance Research, Bellevue, Washington 98004, USA}
\affil[2]{ByteDance Research, Beijing, Beijing 100098, China}

\affil[*]{equal contributions}
\affil[+]{work done as intern at ByteDance Research}
\affil[**]{corresponding: \{weihao.gao, wen.yan\}@bytedance.com}

\begin{document}
\maketitle

\begin{abstract}

Despite the widespread applications of machine learning force fields (MLFF) in solids and small molecules, there is a notable gap in applying MLFF to simulate liquid electrolyte, a critical component of the current commercial lithium-ion battery. In this work, we introduce BAMBOO (\textbf{B}yteDance \textbf{A}I \textbf{M}olecular Simulation \textbf{Boo}ster), a predictive framework for molecular dynamics (MD) simulations, with a demonstration of its capability in the context of liquid electrolyte for lithium batteries. 
We design a physics-inspired graph equivariant transformer architecture as the backbone of BAMBOO to learn from quantum mechanical simulations. Additionally, we introduce an ensemble knowledge distillation approach and apply it to MLFFs to reduce the fluctuation of observations from MD simulations. Finally, we propose a density alignment algorithm to align BAMBOO with experimental measurements. BAMBOO demonstrates state-of-the-art accuracy in predicting key electrolyte properties such as density, viscosity, and ionic conductivity across various solvents and salt combinations. The current model, trained on more than 15 chemical species, achieves the average density error of 0.01 g/cm$^3$ on various compositions compared with experiment. 

\end{abstract}

\section{Introduction}

Liquid electrolyte is an indispensable component in most of the current commercial lithium-ion batteries \cite{xu2004nonaqueous,xu2023electrolytes,meng2022designing}.
The existing commercial electrolytes are mostly carbonate-based, and often composed of more than five components to meet different requirements.
Experimentally exploring molecular interactions for the rational design of electrolytes is expensive, time-consuming, and heavily relied on chemists' intuition and experience. Such limitation poses challenges in transitioning from proof-of-concept experiments in laboratories to products in markets, particularly due to the exponentially increasing complexity involved in optimizing properties of multi-component liquid electrolytes.

Atomistic simulation is an accessible, efficient and flexible alternative to exhaust experimentation. However, to achieve reliable simulation, requirements such as high accuracy and sufficient time-scale and length-scale are necessary.
Quantum mechanical simulation offers high accuracy in describing electronic properties, yet they are computationally intensive and impractical for studying large-scale and complex systems, such as liquid electrolytes. 
On the other hand, classical force fields, while computationally efficient, often sacrifice accuracy in capturing the intricate solvation structures and dynamic behavior of electrolytes. Hence, there exists a pressing need for a balanced and general approach that can reconcile the trade-off between accuracy and speed in modeling liquid electrolytes with different molecular solvents and varying salt types and concentrations. 

In recent years, there has been a growing utilization of machine learning force fields (MLFFs)~\cite{unke2021machine} to perform molecular dynamics (MD) simulations~\cite{frenkel2023understanding}. This trend is primarily attributed to MLFFs' ability to deliver results significantly faster than $ab$-$initio$ quantum mechanical simulations, while fit quantum mechanical data with higher accuracy compared with classical force fields. 
The development of MLFFs has seen two prominent trends. On one hand, there has been a gradual integration of concepts from the field of machine learning into force field design. This evolution has seen a shift from local descriptors-based, rotation-invariant MLFFs~\cite{bp-potential, deeppotential, ace} towards graph neural network (GNN)-based models~\cite{schnet} with rotation-equivariant representations~\cite{nequip} and the transformer architecture~\cite{equiformer, torchmdnet}. On the other hand, there has been an emphasis on incorporating interactions grounded in clear physical foundation into MLFFs. These include electrostatic~\cite{4ghdnnp, dplongrange}, dispersion~\cite{aimnet, physnet}, spin-spin interactions~\cite{spookynet,spindependent}, and so on.

With advancement in the model architecture of MLFFs, the concept of ``universal machine learning force field'', which aims to employ a single MLFF to simulate a wide range of systems, has garnered increasing attention within the realms of solid-state materials~\cite{m3gnet, chgnet, genome, dpa2, macemp} and bio-organic molecules~\cite{aimnet,mace23}. However, when it comes to liquids, particularly liquid electrolytes containing solvents and ions, a universal MLFF that can accurately predict multiple properties across various solvents and salts is still lacking. This limitation may arise from the complex local structures inherent in liquid electrolytes, such as the coexistence of different structural motifs like solvent-separated ion pairs (SSIP), contacted ion pairs (CIP), and aggregates (AGG). As a result, despite some studies utilizing MLFFs to investigate aqueous systems~\cite{mlffwater, liclwater}, molecular liquids~\cite{macesolvent}, and ionic liquids~\cite{mlffil}, there remains a notable scarcity of research specifically focused on MLFFs for liquid electrolytes.

To the best of our knowledge, only two notable previous attempts have been made to simulate liquid electrolytes using MLFFs: Wang $et$ $al.$~\cite{chengjun} utilized Deep Potential~\cite{deeppotential} to calculate the density and solvation structure of LiFSI in triglyme, and Dajnowicz $et$ $al.$~\cite{qrnnliquidelectrolyte} employed charge-recursive neural network (QRNN)~\cite{qrnnmethod} to compute the density, viscosity, and diffusivity of LiPF$_6$ in carbonate solvents. Despite achieving some success, these studies lack conclusive evidence regarding the generalizability of their findings across a wide range of liquid electrolytes. For instance, Ref.~\cite{qrnnliquidelectrolyte} shows that QRNN struggled to achieve high accuracy across both linear and cyclic carbonates simultaneously, as well as solutions with low and high concentrations of LiPF$_6$.

In addition to the scarcity of studies on MLFFs for liquid electrolytes, there are two overarching challenges associated with MLFFs. As highlighted by Fu $et$ $al.$~\cite{fu2022forces}, MD simulations employing MLFFs often encounter issues of collapsed simulations and large fluctuation of results, which stem from the inherent randomness in machine learning~\cite{Ormeño2024} and limit the practical usability of MLFFs. Moreover, most of deep learning-based MLFFs solely rely on learning from quantum mechanical simulations, which do not necessarily guarantee accurate reproduction of experimental measurements across diverse atomistic systems. Although the concept of differentiable molecular simulation~\cite{dmff,pone.0256990} has been introduced to optimize classical force fields based on experimentally measured macroscopic observables, currently there is few deep learning-based MLFFs whose parameters are directly optimized using experimental results. One underlying reason for this absence could be that current optimization methods based on differentiable molecular simulations rely on backpropagating the gradients of MD trajectories to train the parameters. This process is computationally expensive and has the possibility to induce overfitting of deep neural network-based MLFFs to the limited amount of experimental data.

In this work, we introduce the BAMBOO (\textbf{B}yteDance \textbf{A}I \textbf{M}olecular Simulation \textbf{Boo}ster) workflow, specifically designed for constructing MLFFs for MD simulations of organic liquids, with a particular emphasis on liquid electrolytes. The main methodological contributions of this paper are summarized as follows: 1) we propose a MLFF model that integrates a graph equivariant transformer (GET) architecture with physics-based separation of semi-local, electrostatic, and dispersion interactions; 2) we introduce the application of the ensemble knowledge distillation algorithm to lower the fluctuation of results obtained from MLFF-based MD simulations;
3) we propose a physics-inspired density alignment algorithm aiming at aligning MLFF-based MD simulations with experimental data. This approach requires only a minimal amount of experimental data and demonstrates considerable transferability to liquids not initially included in the alignment process. As a result, BAMBOO achieves state-of-the-art accuracy in predicting the density, viscosity, and ionic conductivity of various liquid electrolytes. The high predictive ability of BAMBOO makes it a valuable tool for electrolyte design driven by molecular structure engineering.

\section{Results}

We illustrate the workflow of BAMBOO in Figure~1a. Initially, we sample local atomic environments within liquid electrolytes as gas-phase clusters and subsequently employ DFT to compute their energies, atomic forces, and charges. In this study, aimed at showcasing the broad applicability of BAMBOO, we include diverse molecules and salts in the DFT dataset, as illustrated in Supplementary Figure 1. Notably, our focus encompasses components commonly found in liquid electrolytes utilized in Li-ion batteries, such as cyclic carbonates, linear carbonates, as well as Li$^+$ cations, and PF$_6^-$, FSI$^-$, and TFSI$^-$ anions. Additionally, we incorporate two frequently used organic liquids, ethanol (EO) and acetone (ACT), along with an engineering fluid Novec7000 to showcase the generalizability of the trained model. Subsequently, the quantities calculated by DFT are employed to train a group of GNNs with different random seeds. To lower the fluctuation of results from MD simulations, these independently trained GNNs are fused into a single unified GNN via ensemble knowledge distillation~\cite{asif2019ensemble}. Finally, we employ experimentally measured data, specifically density in this context, to align the MLFF model with experimental observations. Further details about the DFT dataset, training methodology, ensemble knowledge distillation, and density alignment algorithm are provided in the subsequent sections.

In Figure 1b, we describe the separation of semi-local, electrostatic, and dispersion interactions in BAMBOO. Given the significance of long-range electrostatic and dispersion interactions in simulating liquid electrolytes \cite{Chandler1983, Kontogeorgis2018}, we explicitly compute their corresponding energies in BAMBOO. BAMBOO takes as input the types of atoms $\{Z_i\}$ and the 3D coordinates of atoms $\{r_i\}$, and computes energy as follows: 
1) \textbf{Semi-local} energy is modeled by a GNN comprising graph equivariant transformer (GET) layers whose detailed structure will be introduced in the following subsection. GET takes the atom types $\{Z_i\}$ and the relative coordinates $\{\Vec{r}_{ij} = \Vec{r}_i - \Vec{r}_j\}$ of edges $(i, j)$ as input, and outputs atom representations $\{x_i\}$ which encode local environments of atoms. We further input $\{x_{i}\}$ into a multi-layer perceptron (MLP) to obtain the neural network predicted energy $\{E_{i}^{\text{NN}}\}$; 
2) \textbf{Electrostatic} energy is computed as follows. The atom representations $\{x_{i}\}$ are fed into another MLP to predict the atomic partial charge $\{q_{i}\}$. We then compute electrostatic energy $\{E_{i}^{\text{elec}}\}$ based on the predicted partial charge under the framework of charge equilibrium~\cite{Poier2019};
3) \textbf{Dispersion} energy $\{E_{i}^{\text{disp}}\}$ is directly computed based on the DFT-D3(CSO) correction~\cite{2015_Schröder}.

Finally, the total atomic energy ${E_{i}^{\text{total}}}$ is obtained by summing ${E_{i}^{\text{NN}}}$, ${E_{i}^{\text{elec}}}$, and ${E_{i}^{\text{disp}}}$, and the total energy of the system $E$ is computed by summing all atomic energies. Further details of the energy model are provided in the \textbf{Methods} section. Regarding forces, as energy is computed based on relative coordinates $\Vec{r}_{ij}$, rather than absolute coordinates $\Vec{r}_i$, the pairwise force can be naturally defined and computed as $\Vec{f}_{ij} = -\frac{\partial E}{\partial r_{ij}}+\frac{\partial E}{\partial r_{ji}}$, satisfying Newton's Third Law ($\Vec{f}_{ij}=-\Vec{f}_{ji}$) for computing microscopic stress~\cite{TORRESSANCHEZ2016224}. The atomic force is then computed by summing the pairwise forces: $\Vec{f}_{i}=\sum_{j\neq i}\Vec{f}_{ji}$.

\subsection{Architecture of GET Layers}

In Figure 1c, d, and e, we present the the Graph Equivariant Transformer (GET) architecture utilized in BAMBOO, which draws inspiration mainly from the architecture of TorchMD-Net~\cite{torchmdnet}. Beginning with the atom types $\{Z_{i}\}$ and the displacement vectors between atoms $\{\Vec{r}_{ij}\}$, we first initialize the atom scalar representation $x_{i}^{0}$ and the atom vector representation $\Vec{V}_{i}^{0}$, as well as the edge scalar representation $d_{ij}$ and the edge vector representation $\Vec{e}_{ij}$. Subsequently, we feed $x_{i}^{0}$ and $\Vec{V}_{i}^{0}$ into the GET layers to update them iteratively (Figure 1c). As depicted in Figure 1d, within each GET layer, scalar representation $x_{i}^{n}$ undergoes a transformer layer to exchange information with its neighbors within a specified cutoff radius (5 \AA~in this study). Figure 1e illustrates the transformer block designed as attention mechanism~\cite{vaswani2017attention} on edges. Following the transformer block, on the scalar side, we obtain the intermediate atom scalar representation $y_{i}^{n}$. On the vector side, along with $\Vec{e}_{ij}$, the atom vector representation $\Vec{V}_{i}^{n}$ is transformed into an intermediate vector representation $\Vec{u}_{i}^{n}$ and another scalar representation $w_{i}^{n}$. Here, the transformation from vector to scalar is achieved through the inner product operation to maintain rotational invariance. Finally, we combine $y_{i}^{n}$ and $w_{i}^{n}$ to obtain the next-layer atom scalar representation $x_{i}^{n+1}$, and combine $y_{i}^{n}$, $\Vec{u}_{i}^{n}$, and $\Vec{V}_{i}^{n}$ to obtain the next-layer atom vector representation $\Vec{V}_{i}^{n+1}$. In this process, scalar $y_{i}^{n}$ is multiplied to a vector to preserve the equivariance of the vector representation. Overall, within each GET layer, all neighboring atoms exchange information with one another through the transformer, and scalar and vector representations also exchange information with each other. Details of the GET architecture are provided in the \textbf{Methods} section. 

In Figure 2a, b, and c, we elucidate the roles of the equivariant feature, the transformer block, and the prediction of partial charges by ablation studies. For predicting energy and forces from DFT, GET demonstrates superior performance compared with the Graph Equivariant Network (GE) and the Graph Invariant Transformer (GIT). Notably, GE lacks the transformer layer, while GIT does not incorporate equivariance features. This comparison highlights the crucial role of both the equivariant features and the transformer within GET, as evidenced by the smaller errors observed. Furthermore, we observe that equivariance may hold greater importance than the transformer, as GE demonstrates smaller errors than GIT, suggesting avenues for accelerating inference in future applications. 
In addition, we conduct a comparison with a model termed GET-no-charge, which excludes the prediction of partial charges. Interestingly, we observed that GET-no-charge outperforms GET in terms of force errors, but it suffers from larger errors in energy due to the long-range dependence of electrostatic energy (proportional to $1/r$) and the reduced capacity of GNN to capture long-range interactions as opposed to the short-range~\cite{gnnlongrange}. As illustrated in Figure 2c, GET exhibits the smallest error in density from MD simulations compared with GE, GIT, and GET-no-charge, likely attributable to its superior performance in predicting DFT quantities overall. These observations collectively emphasize the suitability of the GET architecture in BAMBOO for simulating liquid electrolytes compared with previous MLFFs without equivariance, transformer, and explicit computation of electrostatic interactions. Further results of the ablation study, along with details of the ablated models, are provided in the Supplementary Table 5.

Compared with TorchMD-Net~\cite{torchmdnet}, our method enhances efficiency through two key strategies. On one hand, we utilize attention mechanisms on graphs, lowering the computational complexity from $\mathcal{O}(N^2)$ to $\mathcal{O}(N)$. On the other hand, we eliminate certain connections and neural network parameters that do not significantly contribute to the capacity of the model. As a result, our model benefits from a noticeable speed enhancement, as illustrated in Figure~2g and h. In this work, we use LAMMPS~\cite{lammps} as the engine to run MD simulations, and we design the interface between BAMBOO and LAMMPS with inspiration from Allegro~\cite{musaelian2023learning}. As shown in Table~\ref{tab:speed_3bpa} and Figure~2g and h, relative to other GNN-based MLFFs like VisNet~\cite{wang2022visnet}, Allegro~\cite{musaelian2023learning}, MACE~\cite{batatia2022mace} or TorchMD-Net 2.0~\cite{palaez2024torch}, BAMBOO achieves a higher inference speed (2 million steps per day for a system with 10,000 atoms on a single NVIDIA A100 GPU). The inference of BAMBOO can be further accelerated by using multiple GPUs in parallel, which will be introduced in a future release.  

\subsection{Ensemble Knowledge Distillation and Density Alignment}

In Figure 1f, we elucidate the concept of ensemble knowledge distillation, which stems from the recognition that the inherent randomness in machine learning can introduce various challenges into MD simulations with MLFFs. Particularly in the case of liquid electrolytes, we observe that GNNs trained with different random seeds, despite exhibiting similar validation errors, may yield divergent macroscopic properties such as density. This discrepancy arises from two main factors: first, MD simulation is inherently a random process~\cite{Ormeño2024}, which in liquid electrolytes can manifest as density fluctuations of approximately 0.004 g/cm$^3$ across different random seeds of MD by the same MLFF; second, during MD simulations, MLFFs often extrapolate to out-of-domain structures, particularly evident in liquid electrolytes where the training data comprises gas phase clusters while the inference domain is bulk liquids. Notably, for neural networks, the degree of randomness is more pronounced for out-of-domain inference compared with in-domain inference~\cite{martius2016extrapolation}, leading to varying behaviors of MLFFs during MD simulations despite being trained on the same dataset.

To address this issue, our approach aims to mitigate the discrepancy among MLFFs in MD simulations by employing an ensemble of MLFFs to predict the energy and forces of MD trajectories. Subsequently, we aggregate the mean predictions and utilize this mean to further optimize the MLFFs. In Figure 2d and Supplementary Figure 7c, we demonstrate that ensemble knowledge distillation effectively reduces the standard deviation of density predictions from five models by more than 50\%, from 0.030 g/cm$^3$ to 0.014 g/cm$^3$. Notably, ensemble knowledge distillation does not require new DFT labels. Beyond BAMBOO and liquid electrolytes, this concept can be applied to any MLFF for any system. In Supplementary Figure 8, we provide another example of utilizing ensemble knowledge distillation to lower the fluctuation of results from M3GNet~\cite{m3gnet} in simulating solid-state phase transformation, further illustrating the generalizability of ensemble knowledge distillation.

As the final step of training BAMBOO, we introduce the concept of density alignment in Figure 1g. For MLFFs targeting liquid electrolytes, we identify two potential sources of systematic bias between MLFF predictions and experimental data. On one hand, the choice of DFT functional, basis set, and dispersion correction may lead to systematic bias on inter-molecular interactions. On the other hand, the deviation between the training data composed of small gas phase clusters and the application scenario of large bulk liquid structures may induce additional bias. 

To address this, we propose aligning BAMBOO with experimental data. Due to the limited availability of experimental data and the high dimensionality of MLFFs, such alignment must be grounded in physics to ensure its transferability. Hence, we employ density as the macroscopic observable for alignment, leveraging pressure as the physics-based link between the macroscopic and microscopic realms. Experimental density can be used to deduce the pressure adjustments required to align MD simulations with experimental density, as depicted in Figure 1g. Moreover, we can correlate these pressure adjustments with inter-molecular forces, followed by utilizing the adjusted force to refine the parameters of BAMBOO.

In Figure 2e, we show that density alignment effectively reduces density errors from around 0.05 g/cm$^3$ to 0.01 g/cm$^3$. Importantly, this reduction in density error, achieved with only 13 experimental data points included in the alignment, transfers to liquids not originally part of the alignment, particularly different solvents and solutions with higher salt concentrations, as demonstrated in Supplementary Figure 10a and Supplementary Figure 12a. 
For example, densities of molecules with bonding characteristics not included in the training set of the alignment, such as VC with the C$=$C double bond, EO with the $-$OH group, and ACT with the ketone group, can be better predicted after alignment. Another notable example is that, the alignment process can improve the prediction of density for electrolytes with salt concentrations (such as 3.70, 3.74, and 3.78 mol/kg) much higher than those in the training set where the highest salt concentration is just 2.22 mol/kg.
This improvement suggests that density alignment, which adjusts the strength of inter-molecular interactions, has some degree of transferability to molecules with similar structures or solutions with varying salt concentrations. Moreover, in Figure 2f, we illustrate that density alignment also decreases errors in predictions of other properties beyond density, such as viscosity and ionic conductivity. Further theoretical analysis, training specifics of the density alignment, and the relationship between density, viscosity, and ionic conductivity are detailed in the \textbf{Methods} section. 

From Figure 2e, 2f and Supplementary Figure 10, it is evident that BAMBOO demonstrates an average density error of 0.01 g/cm$^3$, viscosity deviation of 17\%, and ionic conductivity deviation of 26\% across a diverse range of molecular liquids and solutions with varying salt concentrations. The level of error exhibited by BAMBOO represents the state-of-the-art compared with existing simulation studies as detailed in Supplementary Table 6, 7, and 8. Moreover, the error magnitude of BAMBOO is close to the degree of variation observed in experimental measurements. For instance, density variation from the same research group typically hovers around 0.01 g/cm$^3$, while viscosity and conductivity exhibit variations of approximately 2\% and 1\% \cite{clio_thesis}, respectively. Across different research groups, the viscosity can vary up to approximately 20$\%$ in some cases \cite{hagiyama2008physical,sasaki2005physical,janes2014fluoroethylen, gores2011liquid}, and approximately 5$\%$ for conductivity~\cite{clio_thesis,dave2022autonomous, logan2018study}. In Supplementary Figure 12, we delve into the MD simulated bulk and microscopic properties of solvents and simple electrolytes using BAMBOO, providing comprehensive insights into the predictions. 

Beyond pure solvents and simple electrolytes, we also provide results of BAMBOO simulations on more practical electrolytes with multiple (from 4 to 8) components~\cite{zhushang} in Supplementary Table 9 and 10. Although the density alignment process is mostly based on simple systems such as pure solvents and electrolytes with less than 4 components, for practical multi-component electrolytes BAMBOO achieves similar degree of prediction accuracy for density, viscosity, and ionic conductivity compared with simple systems in Supplementary Figure 10, which further demonstrates the transferability of density alignment in terms of number of components in the system. 
More importantly, we show that for these multi-component systems BAMBOO achieves higher prediction accuracy than the current \textit{off-the-shelf} OPLS-AA classic force field~\cite{oplsaa} for simulating liquid electrolyte, which highlights the potential of BAMBOO on simulating and designing practical liquid electrolytes. 

As for transferability to molecules not included in the DFT dataset, in Supplementary Table 11 we analyze the trend and degree of transferability of BAMBOO to unseen chemical species. We find that BAMBOO can stably simulate unseen molecules with bonding types similar to those included in the DFT training set in Supplementary Figure 1. However, we find that for species with unseen element and unseen bond types (such as C-N triple bond), BAMBOO cannot achieve long-time stable MD simulations. For the molecules that can be stably simulated by BAMBOO, we can see that BAMBOO has the density MAE of 0.0495 g/cm$^{3}$ and the mean viscosity deviation of 52.4\%, which is higher than that of systems included in the DFT training set (0.011 g/cm$^{3}$ and 16.7\%). To further reveal the trend of density prediction on unseen molecules, we conduct MD simulations on fluorinated solvent molecules as in Supplementary Figure 14. We find all MD simulations of the fluorinated solvents (EC, DMC, and EA) remain stable during the MD simulations within a broad temperature range of 283-343 K. For density prediction, we find that the more structural-alike the unseen fluorinated molecule compared with the base molecule, the more accurate the prediction of density from BAMBOO.

To make machine learning force field more transferable to out-of-sample systems, we need constant effort to improve the MLFFs. In general, there are two approaches to elevate the transferability of MLFFs. The first approach is to conduct large scale pretraining on datasets with diverse molecules, such as the MACE-OFF23~\cite{mace23} which is trained on nearly a million conformers. Although the strategy of large scale pretraining is very popular in the recent wave of universal force fields~\cite{m3gnet, chgnet, genome, dpa2, macemp}, it is not proved to guarantee stable MD simulations for all molecules. The other approach is to combine machine learning model with strong physical restrictions, such as our latest work that employs machine learning models to parameterize force fields with classic functional forms~\cite{byteff}, which ensures that the force field can stably simulate any molecule without spurious reaction during the MD simulation. However, such restrictions also limit the usage of machine learning force field on cases where bond breaking happens. Therefore, we think it is important to further develop machine learning force field that can stably simulate as many molecules as possible while preserve the ability to model bond breaking events when necessary. In the very recent work by Niblett $et$ $al.$~\cite{niblett2024transferability}, it is shown that the GNN based model is more transferable to new molecules than local descriptors-based model, which we hope can inspire further development of machine learning architecture for higher transferability of force fields. 

\subsection{Solvation Structures and Atomic Partial Charges}

In addition to accurate predictions of macroscopic properties, BAMBOO's ability to explicitly predict atomic partial charges enables it to provide additional insights about solvation structures in liquid electrolytes. In Figure 3, we present the Li$^+$ and O$_{\text{FSI}^-}$ charge histograms derived from simulations of three distinct LiFSI electrolytes spanning from 1.12m to 3.74m (``m" as molality, mole of salt per kilogram of solvent). The weight ratio of DMC to EC, set at 51:49 wt$\%$, remains consistent across all these electrolytes. The histograms of all the other atoms of FSI$^-$ are presented in Supplementary Figure 15. 

Three distinct peaks can be observed in Figure 3a, each intersecting the X-axis (Li$^+$ charges) at approximately 0.622, 0.626, and 0.632, respectively. As we transit from Figure 3a to Figure 3c, the peak locations roughly persist, albeit with varying heights. Specifically, the highest peak undergoes a shift from approximately 0.632 to 0.622 as the LiFSI concentration increases from 1.12m to 3.74m. 

This shift in Li$^+$-charge distributions is a manifest of the evolving solvation structure populations across different salt concentrations. To demonstrate this interpretation, we analyze the frames of the final 3 ns of MD simulations, and examine solvation structures around Li$^+$s. 
The types of solvation structure and their fractions are tabulated and shown in each Li$^+$ charge histogram. These fractions are derived by first analyzing the radial distribution of the simulation trajectory for the radius of the first solvation shell around Li$^+$, followed by extracting the central Li$^+$ ions along with the surrounding solvated molecules. We find that the first solvation shell has a radius around 2.2 \text{\AA} and the coordination numbers in all three systems are around 4 as shown in Supplementary Figure 16. 

In general, solvation structures are classified as SSIP, CIP, and AGG. Here, SSIP denotes the structure composed solely of solvent molecules, CIP the structure containing one anion, and AGG the collective structure composed of more than one anion in the first solvation shell. Our analysis reveals SSIP ratios of 0.751, 0.494 and 0.182 for LiFSI concentrations of 1.12m, 2.25m, and 3.74m, respectively. CIP ratios are 0.171, 0.260, and 0.276; AGG ratios are 0.078, 0.246 and 0.542. 
The differences in ratio between SSIP, CIP, and AGG in each figure panel and their changes as a function of salt concentrations roughly mirror the peak height evolution depicted in Figures 3a to 3c. 

It is well known that solvation interactions between Li$^+$ ions and surrounding molecules, including solvents and anions, primarily occur through Li$^+$-O interactions in LiFSI-based carbonate electrolytes.
These intermolecular interactions may lead to polarization-induced changes of partial charge within the local solvation environment. To demonstrate this, in addition to analyzing Li$^+$, we also examine the atomic charge distributions across all types of atoms within the FSI$^-$ anion. Interestingly, we observe that O$_{\text{FSI}^-}$ exhibits three overlapping distributions as shown in Figure 3d-f, whereas the other atom types display a single distribution for all concentrations, as presented in Supplementary Figure 15. 

As depicted in Figure 3d, it is evident to identify two peaks (the right and middle), which are highlighted with the red dotted lines. 
We posit these peaks originate from two types of Oxygen atoms within the FSI$^-$ molecule. This hypothesis stems from our observation that at 1.12m LiFSI concentration, most of the anions should remain unpaired due to low CIP and AGG fractions. In addition, these unpaired anions are likely not solvated by solvent molecules. In lithium battery electrolytes, solvents such as EC and DMC are designed to primarily solvate positive ions, resulting in anions often free from forming solvent-anion pairs~\cite{xu2023electrolytes}. Consequently, the two prominent distributions with peak locations around -0.370 and -0.360 are indicative of potential intra-molecular differences in O$_{\text{FSI}^-}$s of free FSI$^-$ in the simulated 1.12m LiFSI electrolyte. 
An additional peak, highlighted with the yellow dotted line in Figure 3d, is visually identified based on its similarity in charge location compared to Figure 3e and 3f. 
We observe this peak progressively becomes more prominent as the concentration increases from 1.12m to 3.74m. We hypothesize that this peak is associated with ion-pairing, as the growth in peak height correlates the increasing total fractions of CIP and AGGs. More discussion about solvation structure and charge prediction from BAMBOO is provided in Supplementary Figure 15 and 16.

\section{Discussions}

In summary, this study presents a machine learning force field framework tailored for MD simulations of liquid electrolytes. First, we devise a GET architecture that segregates semi-local, electrostatic, and dispersion interactions, leveraging knowledge from DFT calculations. Second, we introduce an ensemble knowledge distillation on MLFF, suppressing the fluctuation of results from MD simulations based on MLFF. Moreover, we propose a physics-based alignment approach to reconcile simulated density with experimental data, thus establishing a connection between macroscopic and microscopic scales. Our results demonstrate the effectiveness of the density alignment process in reducing disparities between simulated and experimental outcomes, extending its benefits to properties beyond density.
To further enhance the model's performance in predicting properties beyond density, such as conductivity, simultaneous and direct alignment of these additional properties is essential. Hence, one of our future endeavors will be dedicated to expanding the alignment process to encompass a wider spectrum of properties.

We conduct a comprehensive assessment for the performance of BAMBOO on solvents and liquid electrolytes. Our simulated results demonstrate that a single BAMBOO model can predict densities, viscosities, and ionic conductivities for a broad range of chemistries with high accuracy. The current BAMBOO model is able to simulate up to 15 species in various compositions.
This robustness underscores BAMBOO's value in facilitating design and optimization of practical liquid electrolytes, which may contain up to 10 components.  
In addition to the state-of-the-art accuracy for predicting bulk properties, we also show the quantification of the relationship between solvation structures and atomic partial charges as a function of electrolyte compositions, providing insights for solvation engineering that are unreachable by classical force fields or DFT. 
Compared with current classic (polarizable) force fields, since BAMBOO is a machine learning force field without explicit restriction of functional forms, it has the potential to be extended to other areas beyond simulating bulk properties, such as simulating chemical reactions in liquids~\cite{reactionsinliquids}.
We envision this work as laying the foundation for the development of ``universal machine learning force fields'' capable of accurately simulating properties and behaviors of most organic liquids. 

\section{Methods}
\label{methods}
\subsection{Graph Equivariant Transformer}
Given atomic structures, we first initialize node scalar representation $x_{i}^{0}$ and node vector representation $\Vec{V}_{i}^{0}$ for atom $i$ based on the atom type $z_{i}$, and initialize edge scalar representation $d_{ij}$ and edge vector representation $\Vec{e}_{ij}$ based on the relative vector between atom $i$ and atom $j$ ($\Vec{r}_{ij}$) within a certain cut off radius $r_{\text{cut}}$ (5 \AA\ in this work).
After initialization, we input $x_{i}^{0}$, $\Vec{V}_{i}^{0}$, $d_{ij}$, $\Vec{e}_{ij}$ into the GET layers. In the n$^{th}$ layer, we first input $x_{i}^{n}$ into a LayerNorm~\cite{layernorm} layer, then we input $x_{i}^{n}$, $||r_{ij}||$, and $d_{ij}$ into the classic QKV-transformer shown in Figure~1e. In the transformer block, we compute the attention weight $a_{ij}^{n}$ and the intermediate node scalar representation $y_{i}^{n}$. Furthermore, we build an intermediate vector representation $\Vec{u}_{i}^{ n}$ to incorporate edge vector representation $\Vec{e}_{ij}$ into the GET layer and sum contributions of neighboring atoms to the vector representation. Finally, we output the updated node scalar representation $x_{i}^{n+1}$ and node vector representation $\Vec{V}_{i}^{n+1}$ for the $(n+1)^{th}$ layer by combining all the scalar and vector information on the node level and the edge level. More details of the GET model are provided in Supplementary Equations (1) to (21).

\subsection{Separation of Interactions}
Once we reach the final GET layer, we compute NN-based per atom energy $E_{i}^{\text{NN}}$ and atomic partial charges $q_{i}$ based on $x_{i}^{n}$ by two fully connected networks. Then, we define the electrostatic energy model based on the charge equilibrium theory~\cite{Poier2019}. In this work, we incorporate the charge equilibrium model (Qeq here after) into our MLFF for efficiency. 
We further use a regularization strategy to make our GNN model directly outputs the Qeq charges such that an expensive iterative equilibrium solver per MD timestep is completely removed. In addition, given atomic structures, we can directly compute dispersion interaction ($E_{i}^{\text{disp}}$ and $\Vec{f}_{i}^{\text{ disp}}$) by the D3 correction~\cite{2015_Schröder}. Finally, we combine NN, electrostatic, and dispersion interactions to compute the total energy and force of each atom. More details of the separation of interactions are provided in Supplementary Equations (22) to (45), Supplementary Table 1, and Supplementary Figure 2 and 3.

\subsection{Initial Training of BAMBOO}
Given the design of the GET architecture and the scheme to separate different interactions, we use a loss function combining energy, force, virial, and charge information to train the neural networks in BAMBOO. Specifically, the charge learned by BAMBOO is determined by four aspects: the first is to fit energy, force, and virial together with the contribution from the neural network; the second is to fit the electrostatic potential by fitting the CHELPG charge~\cite{chelpg}; the third is to fit dipole moment, a macroscopic observable derived from charge densities by DFT; and the fourth is to approximate charge equilibrium. 

In this work, we build the neural networks in BAMBOO by PyTorch~\cite{pytorch}, and we use the AdamX optimizer~\cite{kingma2017adam} with the learning rate of $10^{-2}$. In the training process, weight decay is set to 0.001, while the learning rate decreases at a rate of 0.99. The batch size is 128 for both training and validation. The training process is composed of 750 training epochs. The model consists of 3 layers with an embedding dimension of 64. It features 16 attention heads, employs a cutoff radius of 5.0 \AA~for each message passing layer, and includes the fully connected networks with two layers for predicting energy and charge. 

Here, we use 720 thousand clusters with the DFT calculated energy, atomic forces, virial tensor, dipole moment, and partial charge (CHELPG) to train BAMBOO. The calculations are performed utilizing the B3LYP~\cite{b3lyp} functional and the def2-svpd~\cite{svpd} basis set, supplemented by density fitting techniques based on the def2-universal-jkfit auxiliary basis~\cite{weigend2008hartree}. Throughout the computational process, the convergence threshold for the self-consistent field is set to $1.0\times10^{-10}$ a.u., with a maximum allowance of 100 iterations. Here, we use the open-source GPU4PySCF package~\cite{gpu4pyscf} developed by ByteDance Inc. Compared with QChem~\cite{qchem} on 32 CPUs, GPU4PySCF on a single V100 GPU can save up to 97\% computation time and 95\% cost. We randomly choose 90\% data as the training set, and 10\% as the validation set to evaluate the performance of the training process. More details of the initial training of BAMBOO are provided in Supplementary Equation (46), Supplementary Table 2, and Supplementary Figure 4-6.

\subsection{Ensemble Knowledge Distillation for MLFF}
In this study, we first train five models purely by DFT data with different random seeds. Subsequently, we randomly select one model to run 600 ps NPT MD simulations and 400 ps NVT MD simulations for all systems shown in Supplementary Figure 7c. With the obtained trajectories, we employ all five models to predict the energy, forces, and virial tensor of 100 randomly selected frames from the trajectory of each system. Finally, we use the mean predictions as the labels to finetune the randomly selected model by the loss function as in Supplementary Equation (52). More details of the ensemble knowledge distillation of BAMBOO are provided in Supplementary Equations (47-53), Supplementary Table 3 and 4, and Supplementary Figure 7 and 8. 

\subsection{Density Alignment}
We employ the ensemble distilled model as the foundational base model for performing density alignment. To initiate the alignment, we first calculate the compressibility ($\beta$) of each liquid. This is achieved by training the distilled BAMBOO to adapt to uniformly assigned pressure changes ($\Delta P$). Utilizing these varied aligned MLFFs, we conduct 600 ps NPT simulations to gather the resulting densities ($\rho$). Consequently, as depicted in Supplementary Figure 9a-m, we establish a linear correlation between $\Delta \ln(\rho)$ and the assigned $\Delta P$ for each liquid. Leveraging this linearity, we can determine the $\beta$ value for each liquid. This enables us to identify the specific $\Delta P$ that should be incorporated into the base MLFF for each liquid to align the $\Delta \ln(\rho)$ accurately. Finally, we employ the loss function specified in Supplementary Equation (58), applying the unique $\Delta P$ identified for each liquid, to align the base MLFF with experimental data. More details of the density alignment of BAMBOO are provided in Supplementary Equations (54-58), and Supplementary Figure 9-11.

\subsection{Molecular Dynamics}
We employ LAMMPS~\cite{lammps} to conduct the MD simulations for BAMBOO. For each system, an initial step involves energy minimization, which is constrained to a maximum of 1000 iterations and 100,000 evaluations. Following this, the liquid systems are subjected to equilibrium under the NPT ensemble for 1 ns, with a step size maintained at 1 fs. The density values presented in this paper represent the mean value during the final 50 ps of the NPT simulation. Subsequently, we carry out a 4 ns production run under the NVT ensemble to determine the viscosity, diffusivity, and ionic conductivity of the systems. Two approaches are used to calculate ionic conductivities, the Nernst-Einstein method and a method described by Mistry $et$ $al$ \cite{mistry2023relative} that built on Stefan-Maxwell diffusivities. More details of the molecular dynamics are provided in Supplementary Equations (64-68).

\section{Data Availability}
DFT datasets of clusters are provided at \url{https://huggingface.co/datasets/mzl/bamboo}~\cite{data}, and the input parameters and template of atomic structures for LAMMPS MD simulations as well as the final trained, ensemble knowledge distilled, and density aligned model of BAMBOO to reproduce the results in the paper are provided at \url{https://huggingface.co/datasets/mzl/bamboo}~\cite{code}. 

\section{Code Availability}
The source codes, including the GET model, the training module, and the LAMMPS interface for MD simulations, are provided at \url{https://github.com/bytedance/bamboo} and \url{https://zenodo.org/records/14603020}~\cite{code}.

\section{Acknowledgements}
The authors acknowledge insightful discussion on ion transport theory with Aashutosh Mistry, an assistant professor at the Colorado School of Mines. The authors also acknowledge the experimental data points provided by Adarsh Dave, a former PhD student at Carnegie Mellon University, upon request. 

\section{Author Contributions Statement}
Conceptualization: Y.Z., W.Y., W.G., Z.M, Z.Yu, S.G., T.Z., X.H., Z.Yang, Z.W., L.C., X.W., S.S., and L.X.;
Methodology: S.G., Y.Z., Z.M., Z.P., H.W., M.C., W.Y., and W.G.; 
Investigation: S.G., Y.Z., Z.M., Z.P., H.W., M.C., X.H., Z.Yu, W.Y., and W.G.;
Supervision: W.G., W.Y., and L.X.; 
Writing: S.G., Y.Z., Z.M., Z.P., H.W., W.Y., and W.G.

\section{Competing Interests Statement}
All authors are employees of ByteDance Inc. when conducting this research project. ByteDance Inc. holds intellectual property rights pertinent to the research presented herein. Furthermore, the innovations described have resulted in the filing of a patent application in China (Patent Application No. 202311322469.2), which is currently pending.

\section{Tables}

\begin{table}[t]
\caption{Comparison of the inference speed of BAMBOO with other GNN MLFFs on a gas-phase single molecule 3-(bezyloxy)pyridin-2-amine (3BPA), focusing on a fair comparison. BAMBOO is compared with other MLFFs with spherical harmonic order $L=1$, as BAMBOO exclusively utilizes scalar and vector embeddings. The experiments are conducted on a NVIDIA A100 GPU.}
\label{tab:speed_3bpa}
\vskip 0.1in
\begin{center}
\begin{threeparttable}
    \begin{tabular}{lcccc}
    \toprule
    Model & MACE (L=1)~\cite{batatia2022mace} & Allegro (L=1)~\cite{musaelian2023learning} & VisNet (L=1)~\cite{wang2022visnet} & BAMBOO \\
    \midrule
     Speed [ms] & 17.5 & 13.0 & 12.2 & \textbf{6.9} \\
    \bottomrule
    \end{tabular}
\end{threeparttable}
\end{center}
\vskip -0.1in
\end{table}

\section{Figure Legends/Captions}

\subsection{Figure 1}
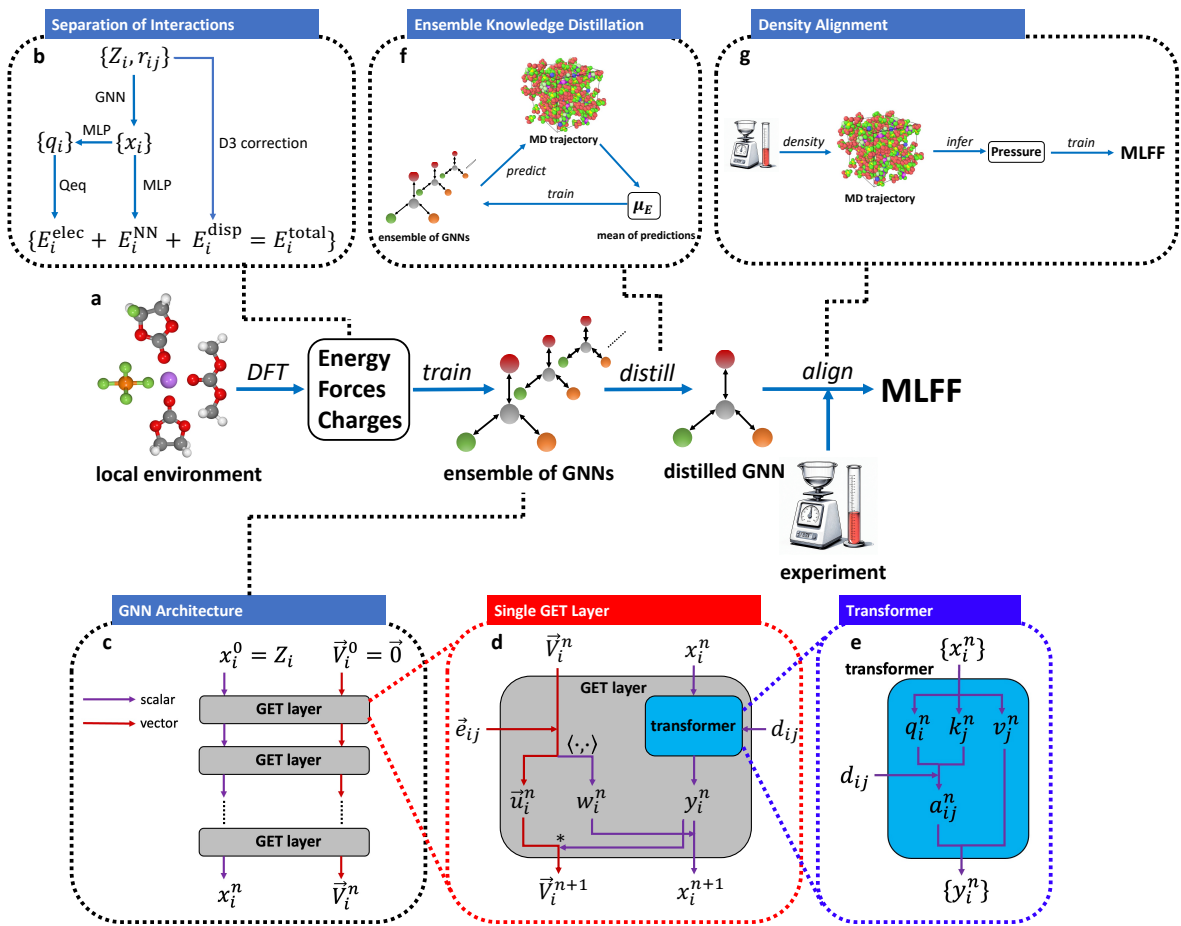
\textbf{Overview of BAMBOO.} \textbf{a.} Schematic of the training process of BAMBOO. \textbf{b.} Separation of interactions in BAMBOO. \textbf{c, d, e.} Schematic of the GNN, the GET layer inside the GNN, and the transformer inside the GET layer, respectively. \textbf{f, g.} Schematic of the ensemble knowledge distillation and density alignment of BAMBOO, respectively.

\subsection{Figure 2}
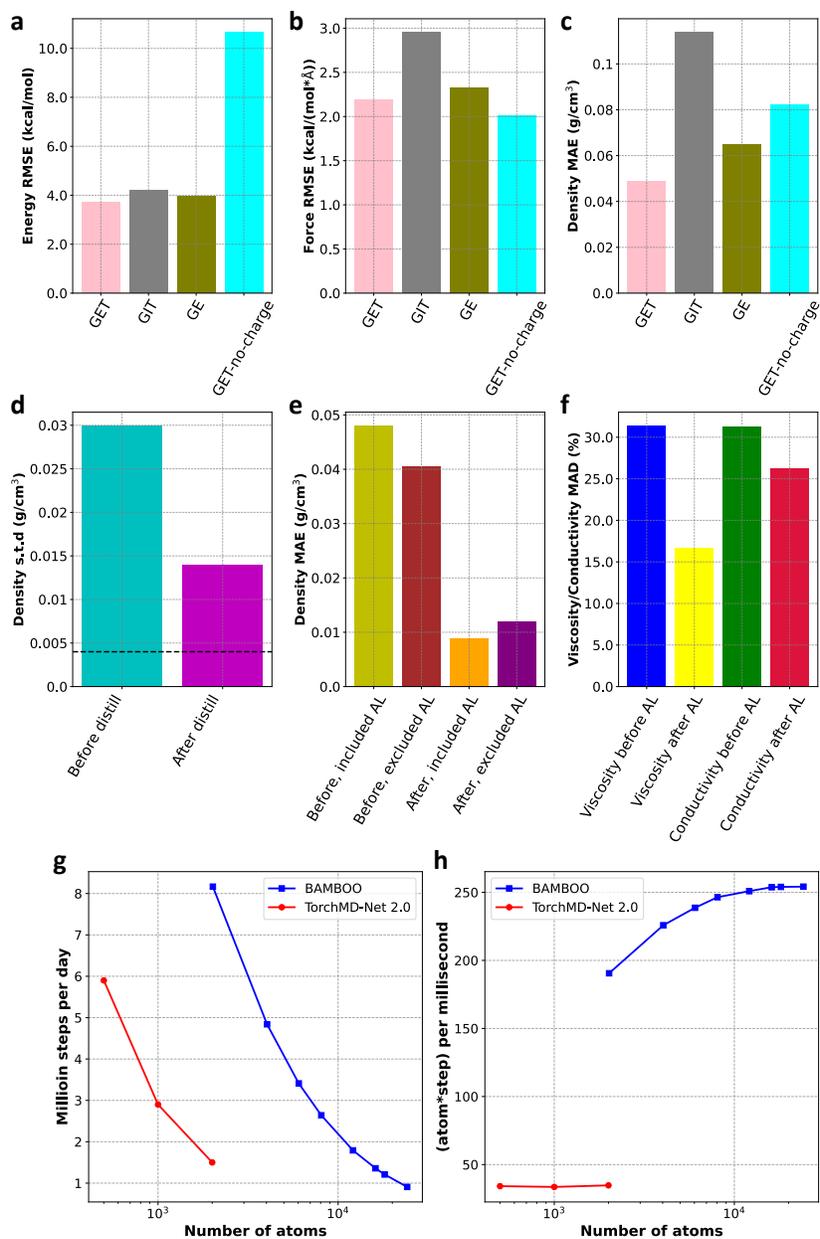
\textbf{Effects of GET layers, ensemble knowledge distillation, and density alignment.} \textbf{a, b, c.} Effects of equivariance, transformer, and charge in BAMBOO. All models are trained using the same DFT training set, excluding ensemble knowledge distillation and density alignment. The evaluation metrics include RMSE (root mean squared error) and MAE (mean absolute error). In the three figures above, bar plots reflect the mean values, and overlaid dots are values from 5 models trained with different random seeds. \textbf{d.} Effect of ensemble knowledge distillation on reducing the standard deviation of density from MD simulations by different randomly trained MLFFs. In this figure, the data are presented as mean values $\pm$ standard deviations of 13 samples in Supplementary Figure 7. \textbf{e, f.} Effect of density alignment on predictions of density, viscosity, and ionic conductivity from MD. The mean absolute deviation (MAD) is used as a metric, expressed in percentage. In the two figures, the data are presented as mean values $\pm$ standard deviations of 13, 15, 27, and 21 samples in Supplementary Figure 10 and Supplementary Table 12-15. \textbf{g, h.} The inference speed comparison between BAMBOO and TorchMD-Net 2.0~\cite{palaez2024torch}. We compare the 3-layer BAMBOO with the 2-layer TensorNet~\cite{tensornet} module implemented in TorchMD-Net 2.0~\cite{palaez2024torch}. We delineate the inference speed in million steps per day in \textbf{g} and in atom times step per millisecond in \textbf{h}. Speeds are tested on a single NVIDIA A100 GPU.

\subsection{Figure 3}
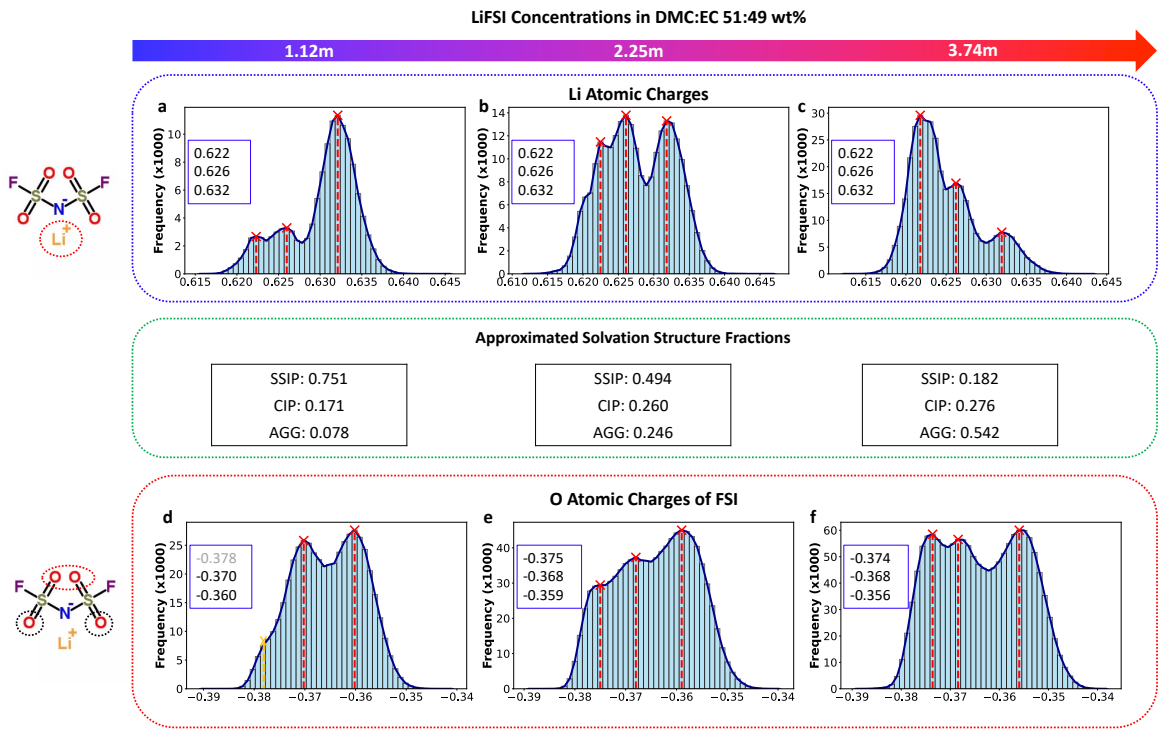
\textbf{The atomic charge distributions and the solvation structure fractions of three simulated LiFSI electrolytes using BAMBOO.}
\textbf{a}, \textbf{b}, and \textbf{c}, arranged from left to right, the Li charge histograms of 1.12m, 2.25m, and 3.74m LiFSI in a DMC:EC mixture of 51:49 wt$\%$. ``m'' denotes molality in mol/kg. Dotted lines within each panel, progressing from left to right, denote the Li charge distributions correlated with SSIP, CIP, and AGG solvation structures. Panels \textbf{d}, \textbf{e}, and \textbf{f}, organized from left to right, the O$_{\text{FSI}^-}$ charge histograms illustrate the behavior within the three LiFSI electrolytes. Notably, the first dotted line may correlates to the distributions of paired O$_{\text{FSI}^-}$. The second and third lines represent the two distinguished O$_{\text{FSI}^-}$ charges linked to distributions of unpaired intra-molecular interactions. The charge values associated with the peaks from left to right are tabulated in legends as shown in each panel.

\bibliographystyle{unsrt}

\end{document}